# Effect of surface mechanical treatment on the oxidation behavior of FeAl-model alloy

Wojciech J. Nowak[1,*] , Daria Serafin[1], and Bartek Wierzba[1]

[1] Department of Materials Science, Faculty of Mechanical Engineering and Aeronautics, Rzeszow University of Technology, Powstancow Warszawy 12, 35-959 Rzeszow, Poland

## ABSTRACT

Fe-based alloys are commonly used in almost every sector of human life. For different reasons, the surfaces of the real parts are prepared using different methods, e.g., mirror-like polishing, grit-blasting, etc. The purpose of the present work is to answer the question how the surface preparation influences the oxidation behavior of Fe-based alloys. To answer this question, a high-purity model alloy, Fe–5 wt%Al, was isothermally oxidized in a thermogravimetrical furnace. The post-exposure analysis included SEM/EDS (WDS) and XRD. The surface roughness was determined by a contact and laser profilometer. The obtained results demonstrate that the mechanical surface preparation influences oxidation kinetics as well as the microstructure of the oxide scale formed on the alloy at both studied temperatures. Namely, polishing and grinding caused local formation of Fe-rich nodules and sub-layer of protective $Al_2O_3$. In contrast, grit-blasting leads to the formation of a thick outer Fe-oxide and internal aluminum nitridation. A significant increase in the oxidation rate of the material after grit-blasting was attributed to grain refinement in the near-surface region, resulting in an increase in easy diffusion paths, namely grain boundaries.

## Introduction

Fe-based alloys are most commonly used as construction materials at both low and high temperatures. The biggest problem in Fe-based alloy usage is their relatively weak corrosion resistance against high-temperature corrosion. However, Fe-based alloys, such as an example steels, are the most attractive materials to be used as construction materials at elevated temperatures due to their relatively low cost. It is well known that low-alloyed steels possess very low high-temperature corrosion resistance [1, 2]. On the other hand, high-temperature Fe-based alloys are used in almost all sectors of human life, e.g., heating elements in toasters, cooking plates, the constructive materials in turbocharger exhaust systems, etc. The shapes of each part used in different systems demand different methods of forming the elements, e.g., cold [3] or hot rolling [4], grit- or sand-blasting [5–7], grinding [8] or even mirror-like polishing [9]. As reported, different surface preparations can significantly change the near-surface microstructure of the material or even result in different level of internal stresses [3–9].

For materials used at elevated temperatures, their oxidation resistance and equally their lifetime at high temperatures become to be a crucial factor during materials engineering. It was found that different surface preparations result in a different oxidation behavior of a wide range of alloys, e.g., Ni-base superalloys [10–15], iron [16], or Fe-base alloys [17–21]. Surface preparation significantly influences the oxidation behavior of these materials. This influence can be positive or negative, depending on the material, exposure condition, chemical composition, etc.

However, most of these researches were performed using commercially available Ni- or Fe-based alloys. Moreover, most of the works about the effect of surface preparation are performed at low temperature (wet corrosion). In most cases, in commercially available alloys, due to several technologically justified reasons, the chemical composition is rather complicated, i.e., they consist of a number of alloying elements. Therefore, it is impossible to unambiguously find out the factor responsible for different oxidation resistances of materials with different surface preparations.

Considering the above-mentioned facts, in the present work, a high-purity Fe–5 at%Al model alloy was investigated to determine the influence of different surface mechanical treatments on oxidation behavior of the investigated alloy. The chemistry of the investigated alloy was chosen due to its ability of forming either a protective $Al_2O_3$ oxide scale or non-protective Fe-oxide scale.

## Materials and methods

In this work, a high-purity Fe-base model alloy, namely Fe–5%Al (given in wt%), has been investigated. From a cylinder with a diameter of 15 mm, 2-mm-thick disks were cut. Each sample had a different surface finishing: polishing up to 1 μm, grinding using 220 grit SiC paper, and grit-blasting. Polishing was performed using a polishing suspension containing $SiO_2$ with a grain size of 1 μm. The samples were grit-blasted using an aluminum oxide powder with grain dimensions approx. 60 mm (220 mesh). The grit-blasting pressure was 0.8 MPa, and the nozzle diameter was 1.5 mm. All of the samples were ultrasonically cleaned in acetone after the preparation and dried by compressed air. Surface roughness of all samples was measured using two methods: contact profilometer HOMMEL Werke T8000 and laser profilometer Sensofar S-Neox Non-contact 3D Optical Profiler with a vertical resolution of 1 nm. The samples with polished and grit-blasted surfaces showed isotropic roughness, while the ground surfaces exhibited anisotropic roughness, namely, the grinding direction can be clearly observed (Figs. 1b, 2b). To exclude the effect of any anisotropy of the ground surface, the roughness measurement by the contact profilometer was always performed in the direction perpendicular to the grinding direction. After roughness measurements, materials were exposed to air at 800 °C and 900 °C for 24 h. After exposure, the samples were electroplated with nickel and mounted in epoxy resin. Metallographic cross sections of the oxidized alloy specimens were prepared by a series of grinding and polishing steps, the final step being fine polishing with the $SiO_2$ suspension with 0.25 μm granulation. The cross sections were analyzed by a light optical microscope (LOM) and scanning electron microscopy (SEM) Hitachi S3400 N equipped with energy-dispersive spectroscopy (EDS) and wavelength-dispersive spectroscopy (WDS) detectors. Phase analyses of the oxidation products of selected samples were performed on selected samples using an X-ray diffractometer Miniflex II made by Rigaku. As the x-ray source, a filtered copper lamp (CuKα, $\lambda$ = 0.1542 nm) with a voltage of 40 kV was used. The $2\theta$ angle range varied between 20° and 120°, and the step size was 0.02°/s. Phase composition was determined using the powder diffraction file (PDF) developed and issued by the ICDD (The International Center for Diffraction Data).

## Results

### Roughness description

Figure 1 shows the SEM/BSE images obtained on polished, ground and grit-blasted surfaces of Fe–5Al. Figure 2 reveals that a more severe surface mechanical treatment results in a higher degree of surface deformation. The results of surface roughness measurement presented in Figs. 1, 2 and 3 undoubtedly show that polished surfaces exhibit the lowest surface roughness, ground surfaces reveal intermediate values of the parameters describing surface roughness, and finally grit-blasting causes the highest surface roughness. Table 1 depicts calculated values of the parameters $R_a$, $R_z$ and $R_{max}$ describing surface roughness. One can see that these parameters differ by approximately an order of magnitude and increase with more severe surface treatments. It should also be noted that an increase in the roughness parameters is directly associated with an increase in the standard deviation.

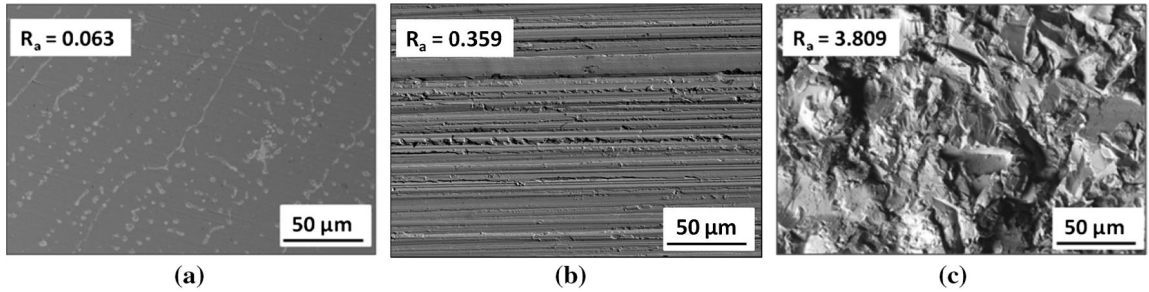

**Figure 1** SEM/BSE images showing topography of Fe–5Al alloy surfaces after: **a** polishing (1 µm), **b** grinding (220 grit SiC paper), and **c** grit-blasting.

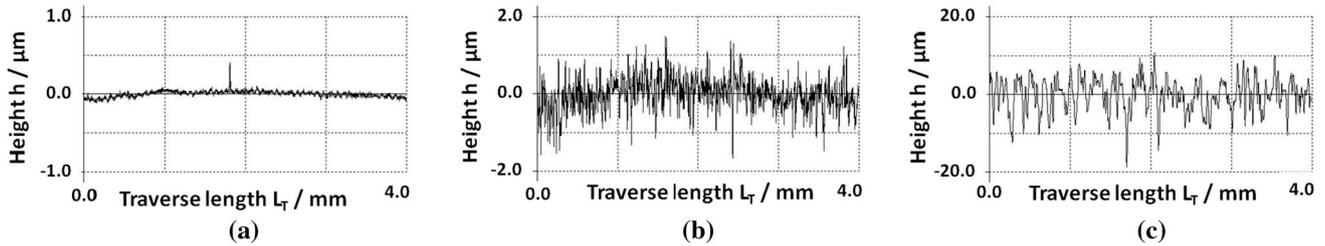

**Figure 2** Surface roughness profiles performed by standard contact profilometer HOMMEL Werk T8000 on **a** polished (1 µm), **b** ground (220 grit) and **c** grit-blasted Fe–5%Al.

Based on a roughness analysis performed using a 3D laser profilometer (the 3D topography of the surfaces is shown in Fig. 3), a parameter Sdr has been calculated for each surface using the equation

**Table 1** Roughness parameters describing samples surface calculated based on measurement shown in Fig. 2

| Parameter | Polishing | | Grinding | | Grit-blasting | |
|---|---|---|---|---|---|---|
| | Average | SD | Average | SD | Average | SD |
| $R_a$ | 0.063 | 0.008 | 0.359 | 0.066 | 3.809 | 0.008 |
| $R_z$ | 0.384 | 0.013 | 3.257 | 0.425 | 24.084 | 0.465 |
| $R_{max}$ | 0.487 | 0.074 | 3.845 | 0.547 | 30.492 | 4.830 |

$$\mathrm{Sdr} = \frac{1}{A}\left[\iint_A \left(\sqrt{\left[1+\left(\frac{\partial z(x,y)}{\partial x}\right)^2+\left(\frac{\partial z(x,y)}{\partial y}\right)^2\right]}-1\right)\mathrm{d}x\mathrm{d}y\right]$$
(1)

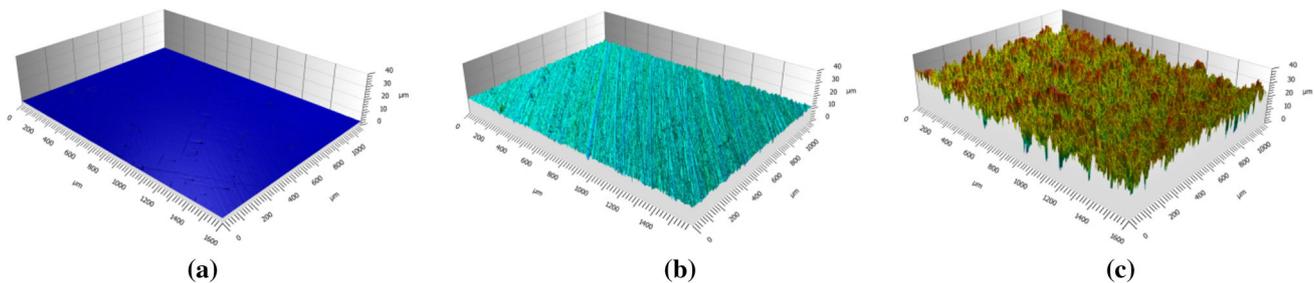

**Figure 3** 3D images of surface topography captured by laser profilometer on: **a** polished (1 µm), **b** ground (220 grit) and **c** grit-blasted Fe–5%Al.

where $A$ is the base area, i.e., the surface area, if the surface were completely flat.

Sdr is the ratio of the difference between the actual and base areas, and the definition area. The developed surface indicates its complexity thanks to the comparison of the curvilinear surface and the support surface. Multiplication of Sdr by 100 allows to express it in %. A completely flat surface will have an Sdr near 0. A complex surface will have an Sdr of higher values [22–24]. The measured value of Sdr for a polished surface is 0.000233, for a ground surface 0.129, and for a grit-blasted surface 0.442. As mentioned previously, the investigated samples were disks with 15 mm diameter and 2 mm height. Then, the calculated base area, assuming completely flat surface, is 4.48 cm$^2$. The real area was calculated using the following equation

$$A_R = A \times (\text{Sdr} + 1) \quad (2)$$

where $A_R$ is real area.

The calculated real areas are equal to 4.58 cm$^2$, 5.05 cm$^2$ and 6.45 cm$^2$ for polished, ground and grit-blasted surfaces, respectively.

## Post-exposure analyses

### Air exposure at 800 °C

Figure 4 shows the obtained mass change of polished, ground and grit-blasted samples during air exposure at 800 °C for 24 h. It is visible from the graph that the samples with the lowest mass gain are that with the ground surface and the mass gain value after 24 h of exposure was equal to 1.10 mg cm$^{-2}$. The sample with a polished surface shows slightly higher mass change as compared to the ground sample, and its value was 1.46 mg cm$^{-2}$. The highest mass change was observed for a sample with a grit-blasted surface, and its value was 4.81 mg cm$^{-2}$. Comparison of mass gains is not a very precise way to compare different types of the materials on high temperature behavior. Therefore, a comparison of the oxidation kinetics using the derived instantaneous apparent oxidation rate constant $K'_w$ is also performed. The procedure for calculating $K'_w$ was described by Quadakkers et al. [25]. Most of the alloys show oxidation according to the classical parabolic oxide scale growth kinetics:

$$\Delta m^2 = K'_w \times t \quad (3)$$

where $\Delta m$ is the specific surface weight change in g cm$^{-2}$, $K'_w$ is the instantaneous apparent oxidation constant, and $t$ is the time in s.

As shown in Fig. 4, after approximately 5 h of exposure, the slope of the mass gain curve decreases for all samples, which means that the oxidation rate slows down. The oxidation kinetics at this stage may be described by the following equation [25]:

$$\Delta m = \Delta m_0 + k_w^{1/2} \times t^{1/2}, \quad (4)$$

where $\Delta m_0$ is an offset value fitted from the transient stage of oxidation.

The instantaneous apparent oxidation constant $K'_w(t)$ is obtained when $K'_w$ is plotted as a function of time with the slope described by the following equation [25]:

$$\left[\frac{d(\Delta m)}{d(t^{1/2})}\right]^2 = K'_w(t) \quad (5)$$

The calculated values of the instantaneous $K'_w$ (Fig. 5) show that the oxidation rate observed on samples with ground and polished surfaces is very similar (1.46 × 10$^{-11}$ g$^2$ cm$^{-4}$ s$^{-1}$ and 2.55 × 10$^{-11}$ g$^2$ cm$^{-4}$ s$^{-1}$ respectively), while the oxidation rate observed for grit-blasted material is significantly higher (2.78 × 10$^{-10}$ g$^2$ cm$^{-4}$ s$^{-1}$).

The SEM/BSE images of surfaces of samples after exposure (Fig. 6) revealed that on the polished surface a relatively smooth and uniform oxide scale was formed. Moreover, crack formation within the oxide scale formed on the polished sample is observed (Fig. 6a). On the ground sample, one can observe that the oxide formation depends on the grinding direction (Fig. 6b). Additionally, in several places, formation of "spiky" shaped oxides is observed. Contrarily, the surface of the grit-blasted material (Fig. 6c) is completely covered by "spiky" shaped oxides.

The SEM/BSE images of the cross sections of polished and ground samples (Figs. 7a, b, respectively) showed similar microstructures, namely local formation of Fe-rich oxide nodules and Al$_2$O$_3$ sub-layers. In contrast, the oxide scale formed on the grit-blasted material revealed a multilayered structure consisting of an outer, dense Fe-rich oxide which forms spikes and sub-layers of porous Fe-rich oxide. Beneath, the presence of internal precipitates of AlN is observed.

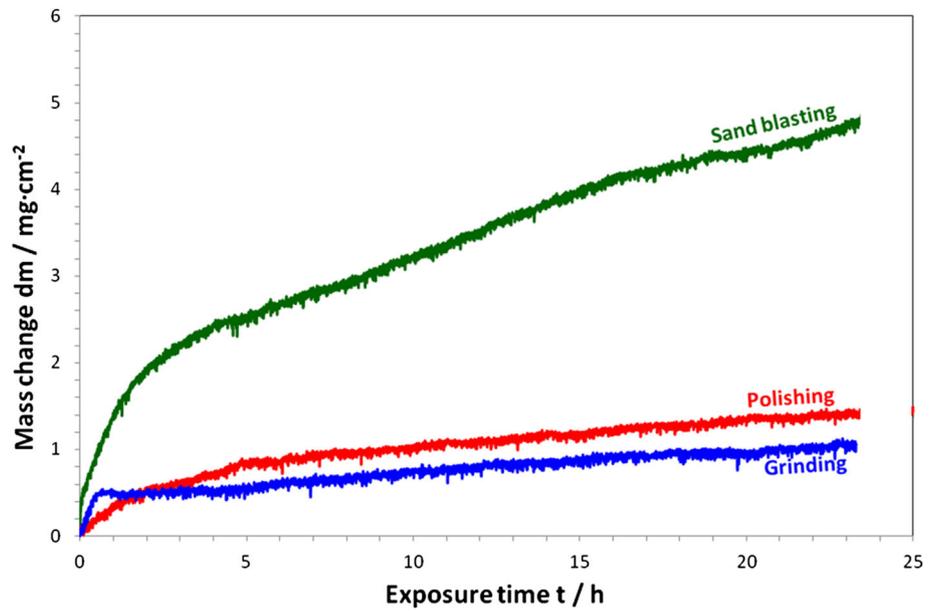

Figure 4 Mass changes obtained for an Fe–5%Al alloy with different surface preparation methods during isothermal air exposure at 800 °C for 24 h.

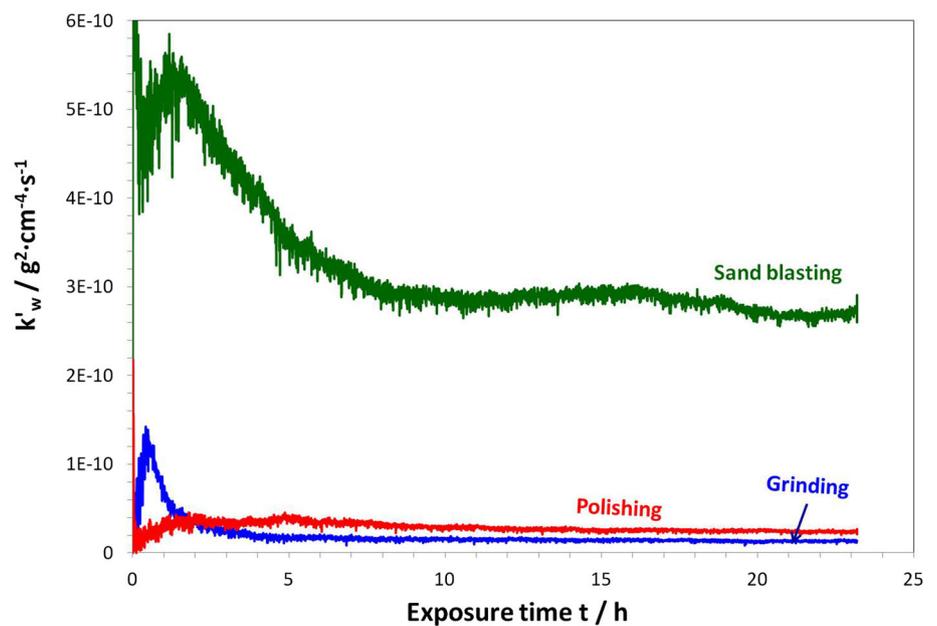

Figure 5 Instantaneous apparent parabolic constant $K'_w$ obtained for an Fe–5%Al alloy with different surface preparation methods during isothermal air exposure at 800 °C for 24 h. The $K'_w$ was calculated using data from Fig. 4 according to the procedure described in Ref. [25].

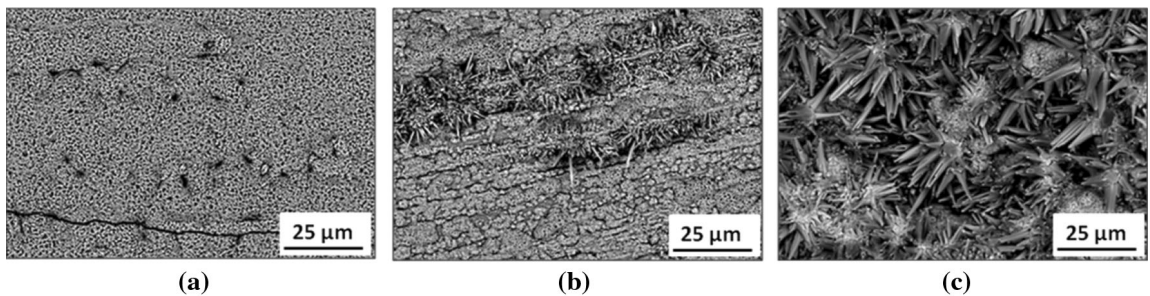

Figure 6 SEM/BSE images showing surfaces of: a polished, b ground and c grit-blasted Fe–5Al after isothermal oxidation test at 800 °C for 24 h in air.

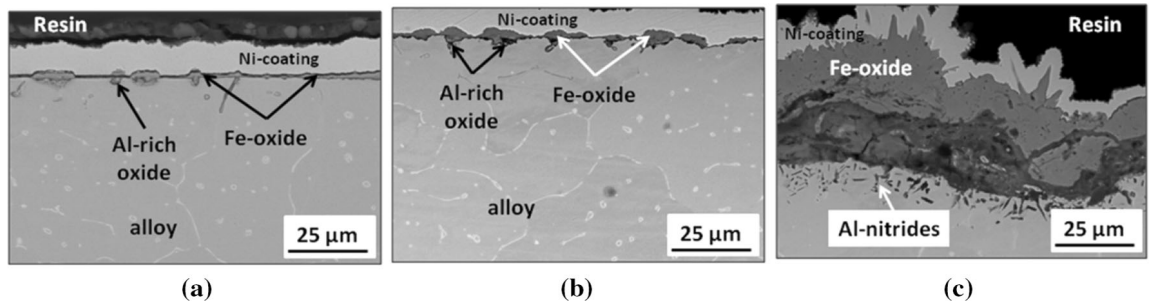

**Figure 7** SEM/BSE images showing cross sections of: **a** polished, **b** ground and **c** grit-blasted Fe–5Al after isothermal oxidation test at 800 °C for 24 h in air.

*Air exposure at 900 °C*

In Fig. 8, the mass changes of polished, ground and grit-blasted samples during air exposure at 900 °C for 24 h are shown. A similar trend as in the case of exposure at 800 °C is observed, namely, the measured values of the mass changes for ground, polished and grit-blasted samples after 24 h of exposure are 0.51 mg cm$^{-2}$, 1.52 mg cm$^{-2}$, and 2.61 mg cm$^{-2}$ respectively. Moreover, the calculated values of the instantaneous $K'_w$ (Fig. 9), show that the alloy with the ground surface revealed the lowest oxidation rate, with polished an intermediate oxidation rate, and grit-blasted the highest oxidation rate (3.43 × 10$^{-12}$ g$^2$ cm$^{-4}$ s$^{-1}$, 2.80 × 10$^{-11}$ g$^2$ cm$^{-4}$ s$^{-1}$, and 8.01 × 10$^{-11}$ g$^2$ cm$^{-4}$ s$^{-1}$ respectively). However, it should be mentioned that the difference between the oxidation rate observed for the samples with polished and ground surfaces and the sample with a grit-blasted surface is not so big as in case of exposure at 800 °C.

The SEM/BSE images of surfaces of samples after exposure (Fig. 10) showed that on polished, ground, and grit-blasted alloys similar oxides are formed as after exposure at 800 °C, namely the ground sample revealed a dependence of the oxides formation on grinding direction (Fig. 10b). As observed after exposure at 800 °C, in several places the formation of "spiky" shaped oxides is observed on the ground surface. Similarly to observation on the grit-blasted sample exposed at 800 °C, after exposure at 900 °C the surface of the sample (Fig. 10c) is also covered by "spiky" shaped oxides (Fig. 11).

The SEM/BSE images of the cross sections of polished, ground and grit-blasted alloys (Figs. 10a–c respectively) showed similar oxide scale

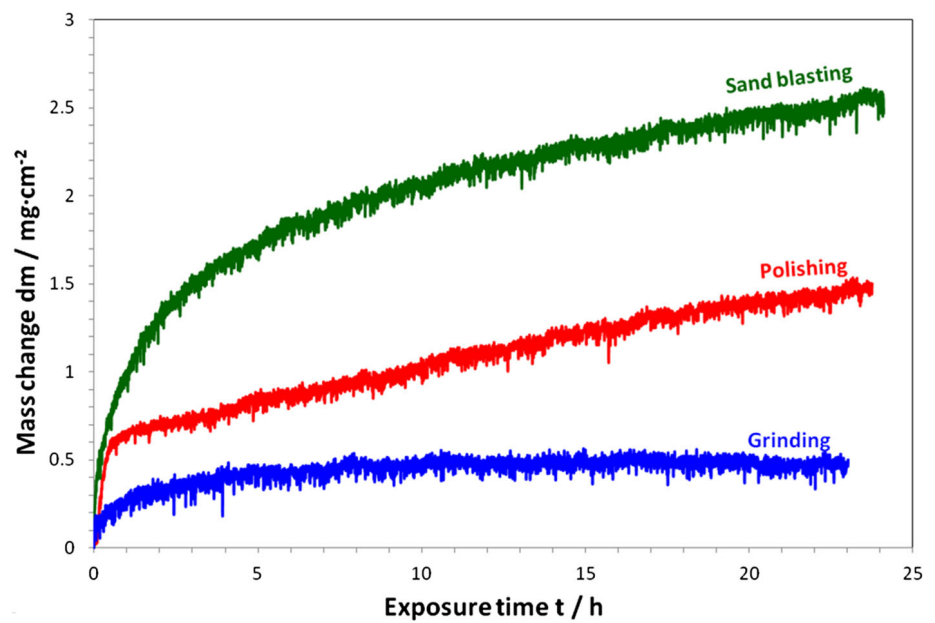

**Figure 8** Mass changes obtained for Fe–5Al alloy with different surface preparation methods during isothermal air exposure at 900 °C for 24 h.

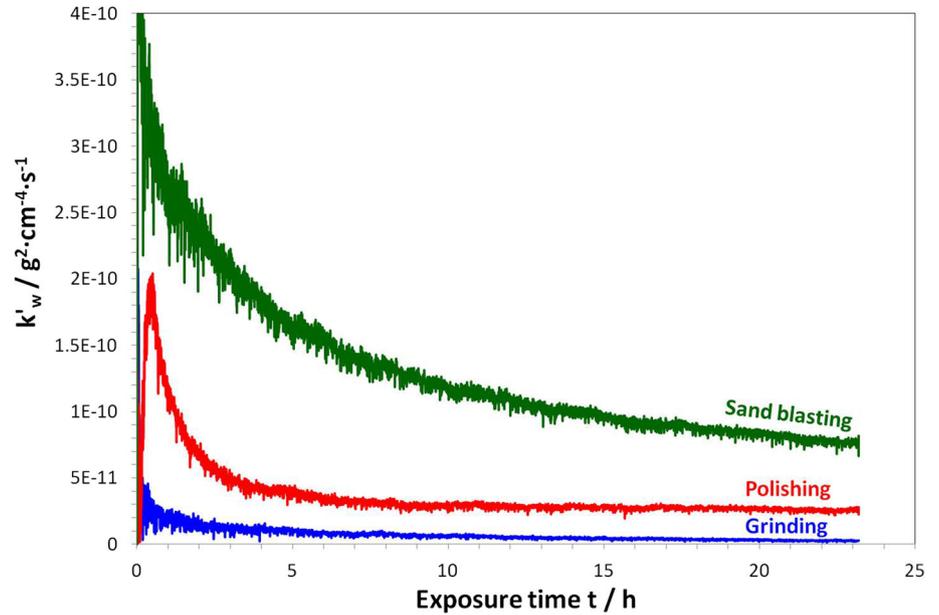

Figure 9 Instantaneous apparent parabolic rate constant $K'_w$ obtained for an Fe–5%Al alloy with different surface preparation methods during isothermal air exposure at 900 °C for 24 h. The $K'_w$ was calculated using data from Fig. 7 according to the procedure described in Ref. [25].

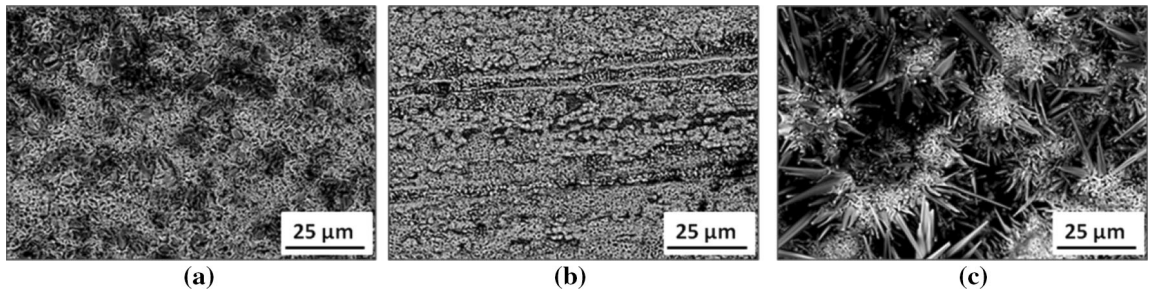

Figure 10 SEM/BSE images showing surfaces of: **a** polished, **b** ground and **c** grit-blasted Fe–5%Al after isothermal oxidation test at 800 °C for 24 h in air.

microstructures as observed after exposure at 800 °C. For polished and ground samples, a local formation of Fe-rich oxide nodules and an $Al_2O_3$ sub-layer are observed (see also Fig. 12), while on the grit-blasted material, a multilayered oxide scale consisting of an outer Fe-rich oxide which forms spikes, below which a zone of internal precipitates of AlN is observed. It should be mentioned that the outer Fe-rich oxide is thinner as compared to the scale formed on the grit-blasted surface after exposure at 800 °C. Moreover, the internal zone of AlN visible in the cross section of alloy exposed at 900 °C is wider comparing with the same material exposed at 800 °C. The zone of the AlN width is 30 μm and 5 μm for exposure at 900 °C and

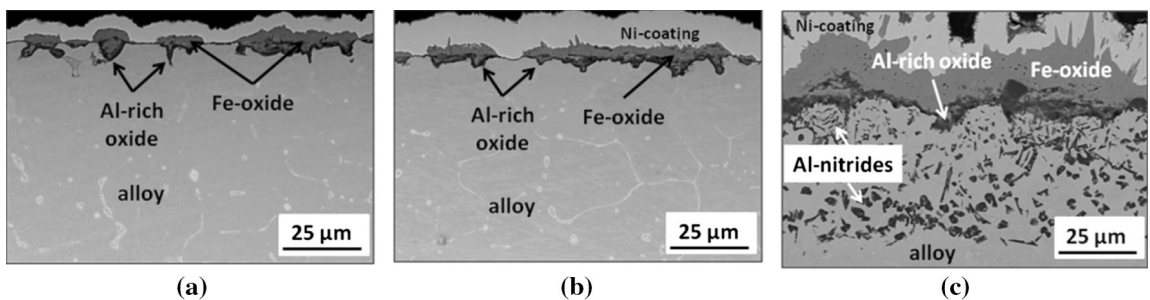

Figure 11 SEM/BSE images showing cross sections of: **a** polished, **b** ground and **c** grit-blasted Fe–5%Al after isothermal oxidation test at 900 °C for 24 h in air.

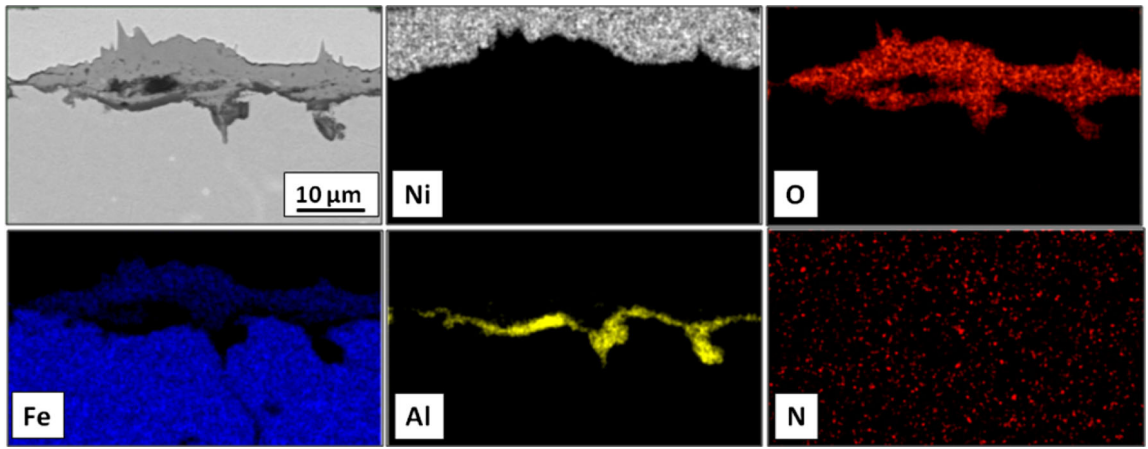

**Figure 12** SEM/BSE image and EDS (WDS for N) elemental maps obtained on the cross section of ground Fe–5%Al after isothermal oxidation test at 900 °C for 24 h in air.

800 °C, respectively. The formation of AlN was confirmed by SEM/WDS elemental maps, as shown in Fig. 13. The latter was also confirmed by an XRD analysis, as shown in Fig. 14. The XRD analysis revealed that the polished and ground samples formed mainly $Fe_2O_3$ (hematite) on the surface, while on the grit-blasted sample apart from $Fe_2O_3$ (hematite), also AlN and $Fe_2O_3$ (high-pressure phase) were found to be formed. The latter phase was previously reported by Bykova et al. [26].

*Exposure at 900 °C in an inert atmosphere*

To elucidate the reason for different oxidation behaviors depending on the surface mechanical preparation, an additional test has been performed. A set of samples with surface prepared by three different methods, namely polished, ground, and grit-blasted, were heat treated at 900 °C for 24 h. To limit the influence of reaction during oxidation, a heat treatment has been performed in an inert atmosphere of high-purity argon. After heat treatment, the surface was slightly polished to reach the near-surface region (roughly 20 μm of material from the surface was removed). After polishing, the samples' surfaces were chemically etched to reveal the grains microstructure. The images of the microstructures after heat treatment show that polishing and grinding did not have an influence on the grain sizes and only primary grains are observed, while on the grit-blasted sample fine secondary grains are visible.

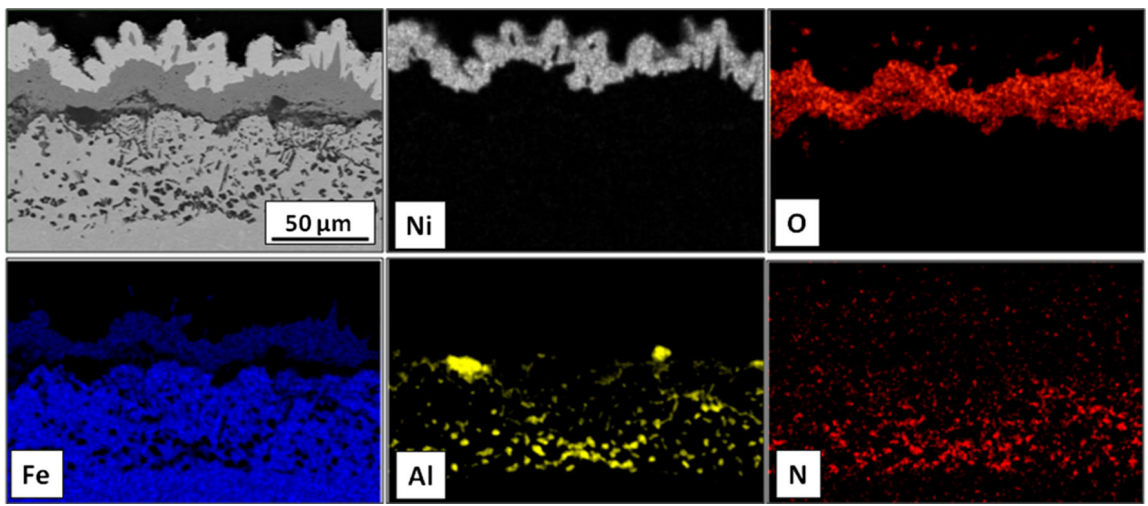

**Figure 13** SEM/BSE image and EDS (WDS for N) elemental maps obtained on the cross section of ground Fe–5%Al after isothermal oxidation test at 900 °C for 24 h in air.

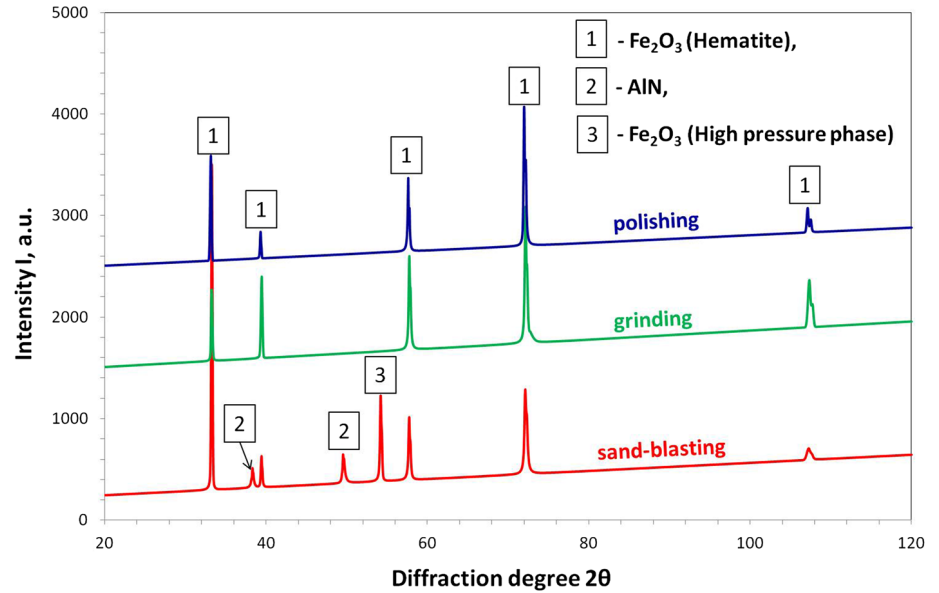

**Figure 14** XRD patterns obtained from polished, ground, and grit-blasted Fe–5%Al after isothermal oxidation at 900 °C for 24 h in air.

## Discussion

The oxidation kinetics measured at both temperatures (800 and 900 °C) revealed clear difference between the samples with differently treated surfaces. At both temperatures, the highest mass gains were obtained for the grit-blasted alloys. The mass change curve of the ground alloy exposed at 800 °C revealed relatively rapid mass gain during the first 30 min of exposure and afterward a very slow increase in mass change. This observation can be explained by the formation of rapidly growing Fe-oxide islands at the beginning of exposure, and a relatively fast $Al_2O_3$ formation which slows down the oxidation rate. The polished material showed a slightly different mass gain curve, namely the mass change curve shows a slightly higher oxidation rate up to 2 h of exposure, and after 2 h the mass change curve becomes parallel to that obtained for the ground surface. The latter means that also polished materials have developed a protective alumina sub-layer; however, the polished material needed a longer time. These observations are confirmed by the SEM images of the cross sections of the samples after exposure. The mass change obtained for grit-blasted alloy was substantially higher as compared to the ground and polished. This means that the alloy mainly formed an Fe-rich oxide during the whole exposure time. The latter was confirmed by the SEM analysis. Moreover, formation of AlN in the form of internal precipitates was observed on the grit-blasted material. Therefore, part of the aluminum was tied up in the internal nitridation zone and its further diffusion toward the oxide scale/alloy interface was hampered. Due to this fact, formation of the protective alumina sub-layer was slowed down and formation of the fast-growing Fe-oxide was enhanced.

A similar trend was observed for the samples exposed at 900 °C. The lowest mass change was observed for the ground alloy, the polished sample exhibited a higher mass change, while the highest mass gain was observed for the grit-blasted alloy. However, one should mention that the mass changes observed for the grit-blasted and ground samples are two times lower compared to exposure at 800 °C, while for the polished material, it is at the same level. It was proposed by Nowak et al. [14] that a rougher surface preparation results in the introduction of a higher number of defects into the near-surface region. These defects are believed to be an easy diffusion path for elements forming the protective oxide scales (aluminum in the present case), which results in a faster formation of the alumina sub-layer, which is indicated by a lower mass change obtained for the ground as compared to the polished surfaces at both studied temperatures. In parallel, similarly to the observation at 800 °C, no nitridation on polished and ground samples was observed. In contrast, a thick zone of AlN below the outer Fe-rich oxide was observed on the grit-blasted sample. However, the thickness of the outer Fe-rich zone is smaller as compared to that formed on the grit-blasted material after exposure at 800 °C. This observation can be explained by two facts. First, as reported by Mehrer [27], the diffusion coefficient for Al in Fe–10 at% Al alloy increases by an order of magnitude from $2.86 \times 10^{-15}$ m$^2$ s$^{-1}$ at 800 °C to $2.74 \times 10^{-14}$ m$^2$ s$^{-1}$ at 900 °C. This results in a thicker AlN zone and the formation of a thin sub-layer of Al-rich oxide at the outer Fe-rich oxide and alloy interface. It is the latter, most probably, which limited the rapid growth of the Fe-rich layer. Formation of Fe-rich, spiky-shaped nodules on Fe–10%Al during oxidation at 900 °C in 1 atm. oxygen was previously reported by Saegussa et al. [28]; however, due to the nitrogen-free atmosphere, formation of AlN below the nodules was obviously not observed. Moreover, no correlation of nodules formation with surface roughness was stated. Usually, if there are no additional factors, Fe–5 wt%Al alloys normally form an external Al$_2$O$_3$ scale accompanied by local Fe-rich nodules, below which an Al$_2$O$_3$ sub-layer is present (as shown in Fig. 12). It has been reported that nitrogen permeability in γ-Fe at 1000 °C is $1.6 \times 10^{-11}$ cm$^2$s$^{-1}$, while permeability for oxygen in γ-Fe at the same temperature is $2.4 \times 10^{-12}$ cm$^2$s$^{-1}$ [29, 30].

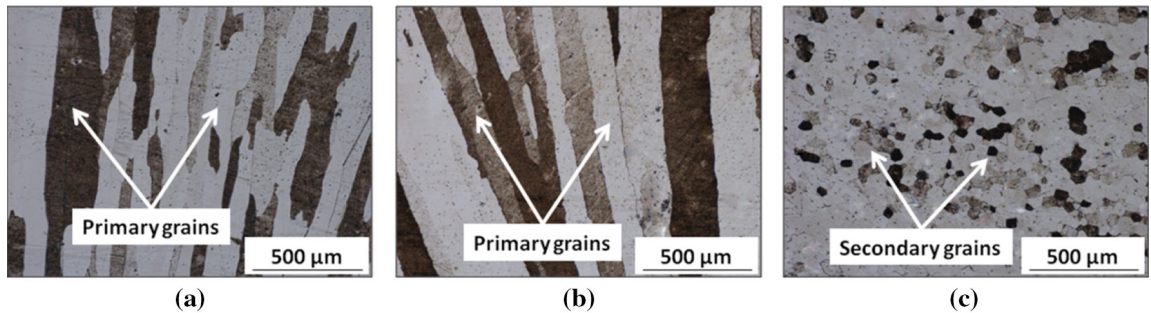

**Figure 15** Images captured using light optical microscope of the near-surface region of chemically etched Fe–5%Al with: **a** polished, **b** ground and **c** grit-blasted surface after isothermal heat treatment at 900 °C for 24 h in argon.

Permeability of the element is defined as [31]:

$$P_X = N_X^{(S)} \times D_X \qquad (6)$$

where $P_X$ is permeability of element $X$, $N_X^{(S)}$ is solubility of the element $X$ in a solid, and $D_X$ is the diffusivity of the element $X$.

However, once a continuous protective Al$_2$O$_3$ layer has been formed, it is a good barrier against nitrogen diffusion; therefore, below the continuous Al$_2$O$_3$ internal precipitates of AlN are not observed. Considering the mass gain curve and cross sections of grit-blasted samples after oxidation, one can conclude that the sample formed non-protective Fe-rich oxides in the form of spikes, which are permeable for nitrogen diffusion and which result in the formation of AlN below the external Fe-rich scale. Formation of AlN was observed by Gurappa et al. [32] after occurrence of "breakaway oxidation" on FeCrAl alloys during exposure at 1000 °C. The AlN precipitates were observed below Fe-rich nodules. Formation of AlN in FeCrAl alloy was observed also by Strehl et al. [33] in crevice, where oxygen partial pressure is too low to form a protective Al$_2$O$_3$ layer. In the case of the alloy studied in the present work, a protective oxide scale was developed, with local nodules of Fe-rich oxides on the polished and ground surfaces. However, as mentioned before, the grit-blasted surface formed an outer Fe-rich non-protective oxide scale and a zone of internal AlN is observed. The latter can be correlated with an increase of 0.442 in the actual area of the grit-blasted surface, which results in more places on which oxygen can adsorb. The latter influences the ratio of oxygen to aluminum atoms present in the near surface region. This in turn enhances the reaction between oxygen and iron, which results in the formation of a non-protective Fe-rich oxide permeable to nitrogen. Moreover, formation of aluminum nitrides and rapid growth of outer spiky-shaped Fe-oxide on the grit-blasted material can be enhanced by recrystallization of the material, which provides easy diffusion paths for nitrogen. As shown in Fig. 15, polishing and grinding did not influence the original material grains shape and size, while grit-blasting caused grains refinement. The latter is obviously accompanied by an increase in the grain boundaries density, which are known as easy diffusion paths for both reagents. Formation of AlN can be attributed to an increase in the inward diffusion of nitrogen, while formation of thick spiky-shaped Fe-oxide is caused by a combination of increased outer diffusion of Fe, and reaction of Al and nitrogen leading to the formation of AlN. The latter substantially hampered the formation of a protective $Al_2O_3$ sub-layer, especially at the lower studied temperature.

## Conclusions

In the present work, the effect of surface mechanical treatment on oxidation behavior of the model alloy Fe–5 wt% Al was studied. The obtained results showed that mechanical surface preparation increases the real specific surface and influences both the oxidation kinetics and the oxide scale microstructures formed on the studied alloy, at both studied temperatures. Namely, polishing and grinding caused local formation of Fe-rich nodules and a sub-layer of protective $Al_2O_3$. In contrast, grit-blasting leads to the formation of thick outer Fe-oxide and internal aluminum nitridation. A significant increase in the oxidation rate of the material after grit-blasting was attributed to grain refinement in the near-surface region resulting in an increase in easy diffusion paths, namely the grain boundaries.

## Acknowledgements

The authors would like to acknowledge Kamil Gancarczyk for performing the XRD analysis.

## Funding

This research was financed within the Marie Curie COFUND scheme and POLONEZ program from the National Science Centre, Poland. POLONEZ Grant No. 2015/19/P/ST8/03995. This project has received funding from the European Union's Horizon 2020 research and innovation programme under the Marie Skłodowska-Curie Grant Agreement No. 665778.